\documentclass[a4paper,cits]{JHEP3c}

\usepackage{epsfig,multicol,bbm,amsmath,amssymb}

\newcommand\fverb{\setbox\pippobox=\hbox\bgroup\verb}
\newcommand\fverbdo{\egroup\medskip\noindent%
            \fbox{\unhbox\pippobox}\ }
\newcommand\fverbit{\egroup\item[\fbox{\unhbox\pippobox}]}
\newbox\pippobox

\def\beq{\begin{equation}}
\def\eeq{\end{equation}}
\def\bea{\begin{eqnarray}}
\def\eea{\end{eqnarray}}
\def\beaa{\begin{eqnarray*}}
\def\eeaa{\end{eqnarray*}}

\def\putunder#1#2{\mathrel{
\setbox0=\hbox{#1}\setbox1=\hbox{\scriptsize #2} \dimen0=-0.5\wd0
\advance\dimen0 by -0.5\wd1 \dimen1=0.5\wd0 \advance\dimen1 by
-0.5\wd1
\hbox{\box0\kern\dimen0%
\vbox to 0pt {\hbox{\lower 0.7em \box1}\vss}%
\kern\dimen1} }}

\newcommand{\gsim}{\lower.7ex\hbox{$\;\stackrel{\textstyle>}{\sim}\;$}}
\newcommand{\lsim}{\lower.7ex\hbox{$\;\stackrel{\textstyle<}{\sim}\;$}}
\newcommand{\be}{\begin{equation}}
\newcommand{\e}{\end{equation}}

\title{Lifetime of Stringy de Sitter  Vacua}

\author{Alexander Westphal\\
    ISAS-SISSA and INFN, Via Beirut 2-4, I-34014 Trieste, Italy\\
    E-mail: \email{westphal@sissa.it}}

\preprint{SISSA-30/2007/EP\\ May 10, 2007}  

\abstract{In this note we perform a synopsis of the life-times from vacuum decay of several de Sitter vacuum constructions in string/M-theory which have a single dS minimum arising from lifting a pre-existing AdS extremum and no other local minima existent after lifting. For these vacua the decay proceeds via a Coleman--De Luccia instanton towards the universal Minkowski minimum at infinite volume. This can be calculated using the thin--wall approximation, provided the cosmological constant of the local dS minimum is tuned sufficiently small. We compare the estimates for the different model classes and find them all stable in the sense of exponentially long life times as long as they have a very small cosmological constant and a scale of supersymmetry breaking $\gsim {\rm TeV}$.}


\begin{document}


\baselineskip=18pt

\section{Introduction}

\baselineskip=18pt

The ongoing search for de Sitter (dS) vacua in string theory -
motivated in part by the recent cosmological data pointing towards
a tiny non-vanishing positive cosmological constant $\Lambda$ - has so far
produced semi-explicit examples firstly~\cite{KKLT} in the context of flux
compactification of the type IIB superstring along the lines
of~\cite{GKP} and, recently, also in the context of M-theory compactified on 7-dimensional manifolds of $G_2$-holonomy without fluxes~\cite{G2a,G2b,G2c} and strongly coupled heterotic M-theory with fluxes on $CY_3\times S^1/Z_2$~\cite{BeCuKr,CuKr}.

The first three examples have in common that the type
IIB axio-dilaton $S$ and the complex structure moduli $U$ of the 6d
compact Calabi-Yau 3-fold are fixed by quantized p-form background
fluxes. These fluxes induce a superpotential for the complex
structure moduli and the dilaton of the Gukov-Wafa-Witten
type~\cite{GVW}.

The fourth example in M-theory does not use fluxes but instead uses a racetrack superpotential generated by non-perturbative effects like membrane instantons or gaugino condensation to stabilize all M-theory moduli in one step~\cite{acharya1,acharya2}.

The fifth example consists of strongly coupled heterotic M-theory with fluxes compactified on $CY_3\times S^1/Z_2$~\cite{BeCuKr,CuKr}. Here, again fluxes stabilize the non-universal moduli, while the Calabi-Yau volume, the orbifold length and the dilaton are stabilized by non-perturbative effects from gaugino condensation, $M2$- and $M5$-brane instantons.

The first class of models - initiated by KKLT~\cite{KKLT} - stabilizes the complex structure moduli and the dilaton with background fluxes. Then they
fix the remaining K\"ahler moduli with non-perturbative effects
like gaugino condensation on $D7$-branes wrapping 4-cycles of the
Calabi-Yau or 4-cycle wrapping Euclidean $D3$-brane instantons.
This produces SUSY anti de Sitter (AdS) vacua for all the moduli.
The uplifting to dS vacua then proceeds:
\begin{itemize}
\item by either introducing an explicitly SUSY breaking
$\overline{D3}$-brane~\cite{KKLT},

\item by supersymmetric D-terms from magnetic world volume fluxes
of 4-cycle wrapping $D7$-branes~\cite{BKQ,Carlos,DuMamb,HaackKrefl,Qbkq2},

\item by the backreaction of $D3$-branes which can provide for a supersymmetric uplifting of a form similar to that of the KKLT $\overline{D3}$-brane~\cite{BurgessD3}

\item by supersymmetric F-terms from the F-terms of hidden sector matter fields~\cite{nilles} (see also~\cite{ReinoScrucca}) or uplifting in (F-term induced) metastable vacua~\cite{DuISS,AHKO,OKKLT1,OKKLT2} along the lines of the ISS-model of SUSY breaking in meta-stable vacua~\cite{ISS}\footnote{Note that the extra field content in these models might introduce additional directions in scalar field space suitable for vacuum decay. The analysis of decay along directions within this new charged field sector would then have to proceed along the lines of the discussion in, e.g.,~\cite{ISS}. When including these models in our analysis we assume tacitly that vacuum decay along these directions has been found to be subdominant compared to the decay in the moduli directions studied here.}

\item or without additional extended sources by taking into
account the effect of the leading higher-order ${\cal
O}(\alpha'^3)$-correction~\cite{bbhl} into the process of
stabilizing the K\"ahler moduli with non-perturbative
effects~\cite{Brama,West,West2}.

\end{itemize} This class of models is characterized by a fine tuning of the flux
superpotential towards small negative values in order to realize
vacua at volumes of ${\cal O}(100)$. The common feature of all the constructions in this class is that the non-perturbative terms in the superpotential determine a lower bound on the width of potential barrier which separates their dS minima from a Minkowski minimum at infinite volume. This is due to the fact that the non-perturbative effects fall off with increasing volume faster than any of the uplifts which have inverse power-law dependence on the volume.

In the second class of models by Balasubramanian, Berglund, Conlon
and Quevedo - the 'large volume scenario' (LVS)~\cite{BBCQ,CQS}
the stabilization of the K\"ahler moduli proceeds via the combined
effects of the leading $\alpha'$-correction~\cite{bbhl} with the
non-perturbative effects which produces non-SUSY AdS vacua at
exponentially large volumes - thus the name. Uplifting then
proceeds either via $\overline{D3}$-branes or D-terms. These
models do not need to have $W_{\rm flux}$ to be tuned small.

The third class of models differs in that the stabilization of the
K\"ahler moduli proceeds solely through the inclusion of
perturbative contributions to the moduli potential. No non-perturbative effects are needed or included. This was done first by compactifying type IIB string theory on Riemann surfaces with closed string fluxes in the presence of $D7$-branes~\cite{Silverstein} (see also~\cite{Silverstein2} for related earlier work in non-critical string theory). This gives, using the fluxes and branes, enough perturbative contributions to the moduli potential to stabilize the complex structure moduli, the dilaton and the K\"ahler moduli altogether in de Sitter vacua. Then, in type IIB flux compactifications along~\cite{GKP}, where closed string fluxes fix the complex structures and the dilaton, the stabilization of the volume can proceed through the inclusion of
perturbative corrections of higher order in $\alpha'$ and the string coupling $g_S$ to the tree level K\"ahler potential:\footnote{For another method stabilizing all closed string moduli perturbatively and at tree level in $\alpha'$ in a Minkowski minimum using both closed and open string fluxes see~\cite{Moral}. However, there the minimum is global and thus no vacuum decay.} Besides the
leading $\alpha'$-correction~\cite{bbhl} there exists a string
1-loop correction~\cite{BHK1} to $K$ which together manage to
stabilize the volume K\"ahler modulus upon a certain tuning of the
flux superpotential at moderately large values in a non-SUSY AdS
vacuum~\cite{GH,BHK2}. Since SUSY is broken in these vacua and the
perturbative nature of the stabilization respects certain shift
symmetries of the underlying string theory, a gauging of these
symmetries via magnetic world volume flux on a 4-cycle wrapping
$D7$-brane then provides an explicit way to uplift these minima to
become dS vacua~\cite{PW}. Note, that for the above two examples in this class the resulting moduli potential for the K\"ahler moduli is qualitatively similar in that it consists of several terms of the form $g_S^r/{\cal V}^s$ ($s\geq 2$) with different signs which thus compete to stabilize the volume ${\cal V}$. Therefore, we will proceed later on to analyze the life-time calculation in the semi-explicit toroidal orientifold example of K\"ahler stabilized dS vacua of~\cite{GH,BHK2,PW}, with the notion in mind that the parametrical life-time estimate obtained there will carry over to the example in~\cite{Silverstein} of type IIB on Riemann surfaces for the above reason.

The fourth class differs significantly in that M-theory compactified on $G_2$-manifolds does not use any background flux. Non-perturbative effects generating a racetrack superpotential are used alone to stabilize all the M-theory moduli while the F-terms of hidden sector charged matter terms allow for a positive vacuum energy. The critical ingredient here is that in M-theory the non-perturbative superpotential generically depends on all moduli~\cite{acharya1,acharya2}. This is different from both the situation of the weakly coupled heterotic string (where racetrack construction to stabilize the dilaton are well studied in the literature) as well as of type IIB string theory where the non-perturbative effects generically depend on the K\"ahler but not on the complex structure moduli.

Finally, the fifth class of models in strongly coupled heterotic M-theory on $CY_3\times S^1/Z_2$~\cite{BeCuKr,CuKr} is in part similar to the KKLT like constructions in that all the non-universal moduli get string scale masses as they are fixed by background flux. The remaining universal moduli, consisting of the dilaton, the $CY_3$-volume and the orbifold length, are then stabilized by a combination of non-perturbative effects alone~\cite{CuKr} which resembles the situation in~\cite{acharya1,acharya2}. The results we get explicitly for the $G_2$-model of~\cite{acharya1,acharya2} will thus carry over to the construction of~\cite{CuKr} because the decay there also proceeds in the decompactificying runaway directions of the universal moduli.

All constructions have in common that they produce a dS vacuum at
tiny positive values $V_0$ of the vacuum energy which is separated
by a high potential barrier $V_1\gg V_0$ from a Minkowski minimum
at infinite volume which corresponds to spontaneous
decompactification of the compact dimensions. These dS vacua
therefore are just metastable under formation of the lower-energy
Minkowski vacuum by quantum mechanical tunneling. This process is
described by the Coleman-De-Luccia (CDL) instanton~\cite{CDL}. In this note we shall then review (for some examples) and calculate (for the other examples) the life--time of the dS vacuum and show that they are all exponentially long--lived.

\section{Metastability of a dS vacuum}

We shall start the discussion thus with summarizing the results of~\cite{CDL} which according to~\cite{KKLT} go as follows. If the potential energy difference $\Delta
V=V_0-V_{\infty}$ of two vacua participating in the tunneling
event is small compared to height of barrier $V_1$ separating
them, i.e. \beq V_1\gg|\Delta V|\;\;,\label{thinwall}\eeq then the thin-wall approximation
becomes applicable. Within this approximation we get the tunneling rate as

\beq \Gamma\sim e^{-\frac{S_{\rm E}(\phi_0)}{(1+4\Delta
V/3T^2)^2}}\label{Gamma}\eeq where $T$ denotes the tension of wall
of the nucleated bubbles of new vacuum

\beq
T=\int_{\phi_0}^{\phi_{\infty}}d\phi\,\sqrt{2\,V(\phi)}\;.\label{tension}\eeq

Here $S_{\rm E}$ denotes the Euclidean action of the scalar field
$\phi$ evaluated at the initial vacuum $\phi_0$. Further, $\phi$ denotes
the canonically normalized direction in scalar field space along
which the tunneling proceeds - i.e. the one with lowest and
thinnest potential barrier. If, as in our cases here, the initial
vacuum is de Sitter with vacuum energy $V_0$ and the final vacuum
is Minkowski, then the expression for the CDL instanton becomes

\beq \Gamma\sim e^{-\frac{S_{\rm
E}(\phi_0)}{(1+4V_0/3T^2)^2}}\;.\label{Gamma2}\eeq

Approximating the bubble wall tension with

\beq T\sim \sqrt{V_1}\,\Delta\phi\label{tension2}\eeq with
$\Delta\phi$ denoting the potential barrier thickness, one arrives
at a universal expression for the decay rate

\beq \Gamma\sim e^{-S_{\rm
E}(\phi_0)}=e^{-\frac{24\pi^2}{V_0}}\label{Gamma3}\eeq as long
as~\cite{KKLT,CDL}

\beq V_0\ll T^2\;\Leftrightarrow
(\Delta\phi)^2\gg\frac{V_0}{V_1}\;.\label{gravcrit}\eeq

Since $V_0\ll V_1$ holds in all the three classes of dS
constructions discussed above it remains to check that the barrier
thickness is not too small (in the above sense): i.e.
$\Delta\phi={\cal O}(0.1\ldots 1)$ would guarantee the longevity
of the dS vacua in all constructions as long as $V_1\gg V_0$ and
$V_0\sim 10^{-120}$. Since

\beq V_1\sim -V_{\rm AdS\;
min.}\sim\left.\frac{|W|^2}{(T+\bar{T})^n}\right|_{T=T_{\rm dS\;
min.}}\;,\;3\leq n<5\label{V1}\eeq in all three model classes
$V_1\gg V_0$ will hold for not too small values of $W$ and the
compact volume ${\cal V}=(T+\bar{T})^{3/2}$ for all of them.

Thus, it remains to check $\Delta\phi={\cal O}(10^{-3}\ldots 1)$ to
establish the longevity of the dS vacua in all constructions.
While this has been done for the KKLT-like
constructions~\cite{KKLT}, which will be recapped in
Section~\ref{KKLTcheck}, to the knowledge of the author this has
not been done for the LVS model~\cite{CQS} and the K\"ahler
stabilization based dS construction~\cite{PW}. The following will
summarize how to determine the barrier thickness $\Delta\phi$ in
these latter constructions. The upshot will be that after identifying the proper canonically normalized field $\phi$ in terms of the K\"ahler modulus $T$ in each type of construction we will find $\Delta\phi={\cal O}(10^{-3}\ldots 1)$ to be valid. This will then establish the longevity of
their dS vacua.

Let us note here that the above estimates hold only for the case that the scalar potential contains just a single dS minimum that has as its final state after the decay only the decompactifying Minkowski vacuum at infinite volume. This can be different if the structure of the scalar potential prior to uplifting is more complicated. Consider as an example a dS vacuum arising from, say, a scalar potential which has two AdS minima of different depth prior to uplifting, where the more shallow AdS minimum shall be at smaller volume. Assume further that SUSY breaking would lift the more shallow AdS minimum to become the fine-tuned small-$\Lambda$ dS minimum as, e.g., in the Kallosh-Linde model~\cite{KL}. Then the decay rate would be enhanced by tens of orders of magnitude compared to the estimate~\eqref{Gamma3} as shown in~\cite{LindeSinks}. 

Thus we shall constrain ourselves here to the study of the simple cases where the dS minimum arises from the uplifting of a scalar potential with a single AdS minimum. This yields afterwards just the dS vacuum and a barrier separating the local minimum from the Minkowski vacuum at infinite volume.

\section{Estimating the barrier width $\Delta\phi$}\label{barrierwidthest}

Let us now outline the method of how to determine the barrier
width $\Delta\phi$ parametrically. In all three classes of models
the 4d ${\cal N}=1$ supergravity AdS scalar potential prior to
uplifting depends on an inverse power of the volume which is
larger than that of the positive semi-definite uplifting
potentials

\beq V_{\rm
AdS}=e^K(K^{T_j\bar{T}_k}D_{T_j}W\overline{D_{T_k}W}-3|W|^2)\sim\left\{
\begin{array}{c}e^{-a_j\cdot T_j}/{\cal V}^2\;,\;\textrm{KKLT}\\ 1/{\cal V}^n\;,\;n\geq
3\;\;\textrm{else}\end{array}\right.\label{Vads}\eeq while \beq
V_{\rm uplift}\sim \frac{1}{{\cal V}^s}\;,\;s\leq
2\;.\label{upl}\eeq

Thus, after uplifting the part of the potential barrier residing
at values ${\rm Re}\,T>{\rm Re}\,T_1$ will be wider than the other
part of the barrier situated between the dS minimum $V_0$ at $T_0$
and the barrier maximum $V_1$ at $T_1$.

For instance, in the KKLT construction (assuming just one K\"ahler
modulus for simplicity) we have

\bea K&=&-3\ln(T+\bar{T})\;\;,\;T=t+i\,b\nonumber\\ \label{KWkklt}\\
W&=&W_0+A\,e^{-a\cdot T}\;,\;W_0=W_{\rm flux}\nonumber \eea
combined with an uplift from an $\overline{D3}$-brane which yields

\beq
V(T)=e^K(K^{T\bar{T}}D_TW\overline{D_TW}-3|W|^2)+\frac{D}{({\rm
Re}\,T)^2}\;\;.\label{KKLTpot}\eeq Here $b$ denotes the 4d scalar
axion partner of the 4-cycle volume measured by $t$.

Fig.~\ref{Fig.1} shows the situation which exemplifies the
asymmetric distribution of the barrier width around the barrier
maximum described above.

\FIGURE[ht]{\epsfig{file=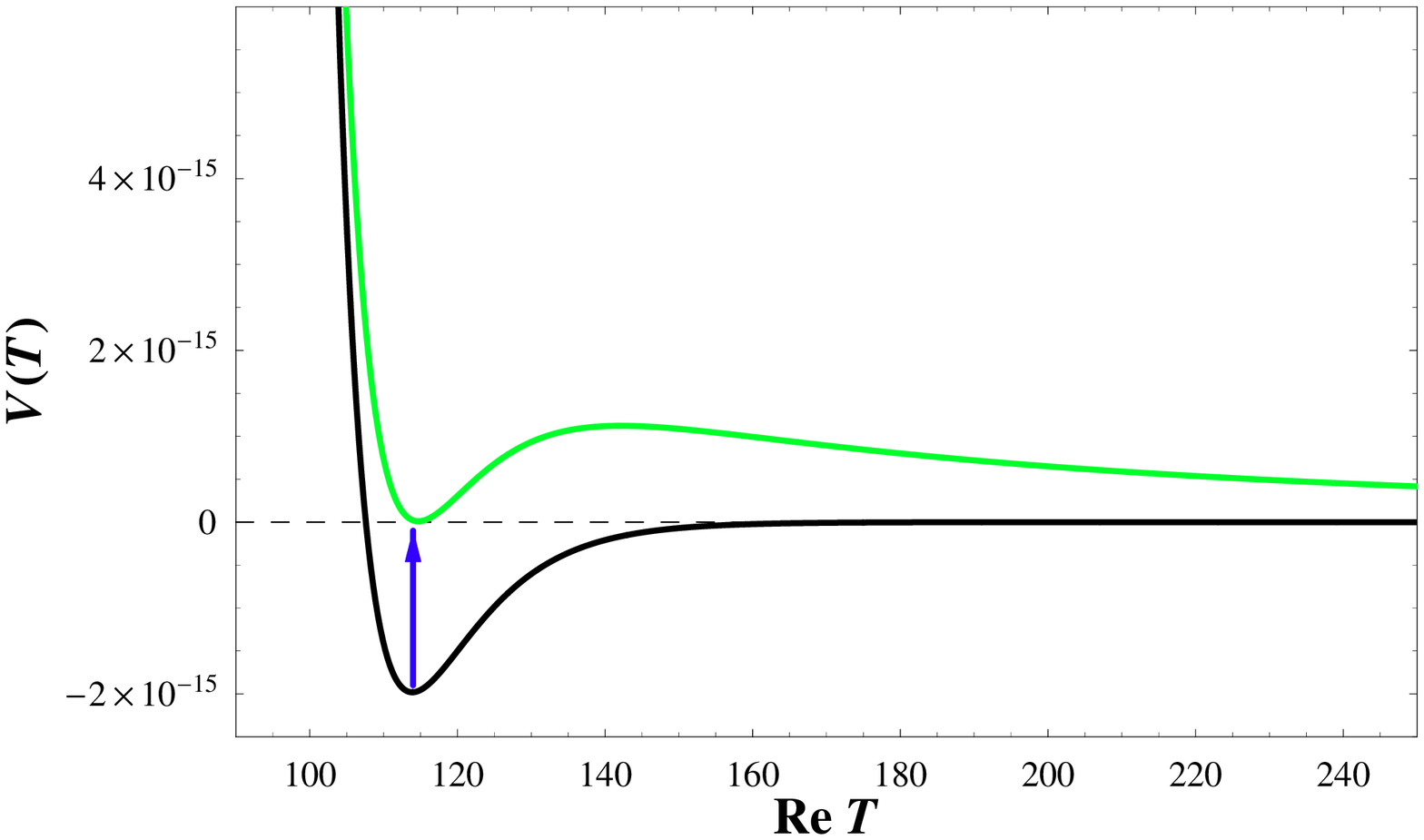,width=12cm} \caption{Black: The F-term scalar
potential $V_F(T)$ leading to the stabilization of $T$. Green: The
scalar potential eq.~\eqref{KKLTpot} after uplifting by switching
on a single $\overline{D3}$-brane.}%
\label{Fig.1}}

In view of this asymmetric barrier width, a conservative estimate
of the width should be given by taking

\beq \Delta {\rm Re}\,T\sim 2\cdot({\rm Re}\,T_1-{\rm
Re}\,T_0)\label{width}\eeq where as before $T_0$ denotes the
position of the dS minimum and $T_1$ the one of the barrier
maximum. Further, an estimate of ${\rm Re}\,T_1$ follows from the
fact, that the barrier maximum arises at that point where the AdS
scalar potential prior to uplifting has decreased its magnitude
significantly compared to its value at the AdS minimum when going
towards larger ${\rm Re}\,T$. We can thus take

\beq {\rm Re}\,T_1:\;|V_{\rm AdS}(T_1)|=\epsilon\cdot|V_{\rm
AdS}(T_0)|\;{\rm with}\;\epsilon={\cal O}(0.1)\;{\rm and}\;{\rm
Re}\,T_1>{\rm Re}\,T_0\;\;.\label{T1}\eeq Using this in
eq.~\eqref{width} should then provide us with a conservative
parametrical estimate of the barrier width in terms of the field
$T$. Next, note that the field $T$ is not canonically normalized.
However, a look at its K\"ahler potential $K=-3\ln(T+\bar{T})$
allows as to define the canonically normalized field we need for
calculating the barrier width used in eq.s~\eqref{Gamma2},
\eqref{tension2} as~\cite{KKLT}

\beq \phi=\sqrt{\frac{3}{2}}\cdot\ln({\rm
Re}\,T)\;.\label{Tcan}\eeq From here we arrive finally at a
parametrical expression for the barrier width $\Delta\phi$ given
by

\beq \Delta\phi\sim\sqrt{6}\cdot\ln\left(\frac{{\rm Re}\,T_1}{{\rm
Re}\,T_0}\right)\;\;.\label{width2}\eeq In the following Chapter
we will apply this formalism to the three classes of models
discussed above.

\section{Comparing the existing constructions}

\subsection{KKLT-like dS constructions}\label{KKLTcheck}

The KKLT construction is described in 4d by the ${\cal N}=1$
chiral ungauged supergravity specified by the K\"ahler potential
and superpotential of eq.~\eqref{KWkklt}. According to the method
outlined in the last Section we determine the barrier width of the
positive semi-definite scalar potential after uplifting by
extracting the position $T_0$ of the AdS vacuum prior to uplifting
(which is close to the position of the later dS minimum) and the
position $T_1$ of the barrier maximum. The former is given by the
solution to

\beq \left.D_TW\right|_{T_0}=-aA\,e^{-a
T_0}-\frac{3\,(W_0+A\,e^{-a
T_0})}{T_0+\bar{T}_0}=0\;.\label{T0kklt}\eeq

Now from here we can infer the position $T_1$ of barrier maximum
directly: At $T_1=T_0+1/a$ the exponential contribution in $W$ is
down by a factor of $1/e$ which implies that

\beq
\left.D_TW\right|_{T_1=T_0+1/a}\approx-\frac{3\,W_0}{T_1+\bar{T}_1}
\;.\label{T1kklt}\eeq However, this expression is the no-scale
result which implies that at this value of $T_1$ we have
$K^{T\bar{T}}D_TW\overline{D_TW}\approx 3|W|^2$ and thus $V_{\rm
AdS}(T_1)\approx 0$. Using eq.~\eqref{width2} this gives us the
barrier width in terms of canonically normalized field $\phi$ as

\beq \Delta\phi_{\rm KKLT}\sim
\sqrt{6}\cdot\ln\left(\frac{t_0+1/a}{t_0}\right)=\frac{\sqrt{6}}{a\,t_0}+{\cal
O}\left(\frac{1}{(a\,t_0)^2}\right)\;.\label{KKLTwidth}\eeq
Validity of the supergravity approximation requires $t_0\gg 1$ and
$a\,t_0\gg 1$ - typical model constructions have $t_0={\cal
O}(100)$ and $a\,t_0={\cal O}(10)$. Thus, in KKLT--like models we
have typically \beq (\Delta\phi_{\rm KKLT})^2=0.01\ldots
0.1\gg\frac{V_0}{V_1}\label{KKLTres}\eeq in accordance with the
requirement of eq.~\eqref{gravcrit}. This in turn implies that the
lifetime of the dS vacuum is given from eq.~\eqref{Gamma3}
as~\cite{KKLT} \beq \tau_{\rm KKLT}\sim\Gamma_{\rm KKLT}^{-1}\sim
e^{\frac{24\pi^2}{V_0}}\,t_P\sim e^{10^{120}}\,t_P\;\;{\rm
for}\;V_0\sim 10^{-120}\eeq where $t_P\sim 10^{-43}\,s$ denotes
the Planck time. Note that these estimates apply directly also to all the other KKLT-like de Sitter vacua constructions in this first class of models (see Section 1.1) along the lines of~\cite{BKQ,Carlos,DuMamb,HaackKrefl, BurgessD3,DuISS,AHKO,OKKLT1,OKKLT2, ReinoScrucca,Brama,West,West2}. This derives from the fact that in all constructions their respective F--term or D--term uplifting produces a $\delta V_{\rm uplift}=C/(T+\bar{T})^n$ which imitates the effect of the $\overline{D3}$-brane of KKLT.

\subsection{LVS type large-volume dS
constructions}

This class of models~\cite{BBCQ} is based on a compactification of
the type IIB superstring on the orientifolded
$\mathbb{P}_{[1,1,1,6,9]}^4$ del Pezzo surface. This manifold has
$h^{1,1}=2$ K\"ahler moduli and $h^{2,1}=272$ complex structure
moduli which latter ones can be stabilized by turning on the
closed string background fluxes. The remaining two K\"ahler
moduli, called $T_4$ and $_5$ in what follows, are then fixed by
introducing non-perturbative superpotential contributions coupling
to each of them and the inclusion of leading $\alpha'$-correction
to the K\"ahler potential. The model is then described in 4d by a
K\"ahler potential and a superpotential as follows

\bea K&=&-2\ln{\cal V}-
\frac{C_{\alpha'}}{\cal V}+\ldots\;,\;{\cal V}=
\frac{1}{9\sqrt{2}}\,(t_5^{3/2}-t_4^{3/2})\nonumber\\ \label{KWcqs}\\
W&=&W_0+A_4\,e^{-a_4\cdot T_4}+A_5\,e^{-a_5\cdot T_5}\nonumber
\eea The constant $C_{\alpha'}$ can be computed in this model and
is given by~\cite{bbhl,BBCQ,CQS}

\beq
C_{\alpha'}=\underbrace{-\frac{1}{2}\,\zeta(3)\chi}_{\xi}\cdot
({\rm Re}\,S)^{3/2}\;\;{\rm
with:}\;\chi=2\,(h^{1,1}-h^{2,1})=-540\;.\label{Calpha}\eeq

The corresponding F-term scalar potential $V_{\rm AdS}$ from
eq.~\eqref{Vads} takes in the region where $t_5\gg t_4>1$ the
approximate form

\beq V_{\rm AdS}=\frac{|W_0|^2}{{\cal V}^3}\,(-\mu\ln{\cal
V}+\lambda\sqrt{\ln{\cal V}}+\nu\,\xi)\;\;.\label{Vcqs}\eeq This
potential fixes the K\"ahler moduli in a SUSY breaking AdS minimum
at $(t_{5,0},t_{4,0})$ such that $t_ {5,0}\gg
t_{4,0}>1$~\cite{BBCQ,CQS}. Because of this regime we have ${\cal
V}=t_5^{3/2}/9\sqrt{2}$ to a very good accuracy at this AdS
minimum implying that tunneling to the decompactifying Minkowski
minimum at infinity will occur practically completely in the
$t_5$-direction in scalar field space. The potential approaches
zero from below beyond the AdS minimum and thus can be
approximated there as \beq V_{\rm AdS}\sim
-\,|W_0|^2\cdot\frac{\mu\ln{\cal V}}{{\cal V}^3}\sim
-\,\frac{\mu\,|W_0|^2}{972\sqrt{2}}\cdot\frac{\ln
t_5}{t_5^{9/2}}\;\;{\rm for}\;t_5>t_{5,0}\;.\label{Vcqs2}\eeq Now
we can apply eq.~\eqref{T1} to extract $t_{5,1}$, i.e. the
position where $|V_{\rm AdS}(T_{5,1})|\ll |V_{\rm AdS}(T_{5,0})|$,
as \beq t_{5,1}=t_{5,0}\cdot\left[1+\frac{2}{9}\gamma+{\cal
O}(\gamma^2)\right]\;\;{\rm
where}\;\gamma=1-\epsilon\;\;.\label{T1cqs}\eeq This allows us to
compute the barrier width which the dS vacuum resulting from
uplifting the above AdS minimum with, e.g., a D-term will have in
terms of the canonically normalized 'tunneling' field $\phi$ from
eq.~\eqref{width2} to yield \bea \Delta\phi_{\rm LVS}&\sim&
\sqrt{6}\cdot\ln\hspace{-0.2ex}\left(\frac{t_{5,1}}{t_{5,0}}\right)\sim
\left(\frac{2}{3}\right)^{3/2}\cdot(1-\epsilon)=0.05\ldots0.2\;\;{\rm
for}\;\epsilon=0.1\ldots0.2\nonumber\\
&\phantom{\sim}&\;\Rightarrow (\Delta\phi_{\rm LVS})^2\sim10^{-3}\ldots 10^{-2}\gg\frac{V_0}{V_1} \label{CQSwidth}\eea This result is
parametrically similar to the KKLT-like models and thus a similar
estimate results for lifetime of the dS vacua in these
large-volume vacua \beq \tau_{\rm LVS}\sim\Gamma_{\rm
LVS}^{-1}\sim e^{\frac{24\pi^2}{V_0}}\,t_P\sim
e^{10^{120}}\,t_P\;\;{\rm for}\;V_0\sim 10^{-120}\;\;.\eeq
We note finally, that these results are stable under the addition of further string loop corrections, as the whole LVS construction has recently been shown to be stable when taking into account these corrections~\cite{LVSloops}.

\subsection{D-term uplifted K\"ahler stabilization dS construction}

This example of the third class of models arises from the observation that the
combined effects of the leading perturbative $\alpha'$-correction
and 1-loop correction to the K\"ahler potential suffice in
presence of the flux superpotential to stabilize the volume
K\"ahler modulus~\cite{GH}. No non-perturbative effects are needed
here. 

Notice, that that the ensuing discussion will carry over qualitatively to the other example in this class~\cite{Silverstein} (type IIB on Riemann surfaces). This is due to the fact that the resulting moduli potential for the K\"ahler moduli is qualitatively similar for both examples in that it consists of several terms of the form $g_S^r/{\cal V}^s$ ($s\geq 2$) with different signs which thus compete to stabilize the volume ${\cal V}$. Therefore, we choose to analyze the life-time calculation in the semi-explicit toroidal orientifold example of K\"ahler stabilized dS vacua of~\cite{GH,BHK2,PW} summarized below for the sake of explicitness. For the above reason then the parametrical life-time estimate obtained there will carry over to the example in~\cite{Silverstein}of type IIB  on Riemann surfaces.

The stringy 1-loop correction has been calculated in a few
explicit orientifold cases~\cite{BHK1} which thus leads to an
explicit realization of this K\"ahler stabilization
model~\cite{BHK2}. The model is described in 4d by a K\"ahler and
superpotential as follows \bea K&=&-3\ln(T+\bar{T})-
\frac{C_{\alpha'}}{(T+\bar{T})^{3/2}}-\frac{C_{\rm 1-loop}}{(T+\bar{T})^{2}}+\ldots\nonumber\\ \label{KWmydS}\\
W&=&W_0\nonumber \eea where $C_{\alpha'}$ is given by
eq.~\eqref{Calpha} and $C_{\rm 1-loop}\sim (U+\bar{U})^2$ with $U$
denoting collectively the complex structure moduli. The resulting
scalar potential \beq V_{\rm
AdS}=\frac{|W_0|^2}{(T+\bar{T})^3}\cdot\left(\frac{A}{(T+\bar{T})^{3/2}}
+\frac{B}{(T+\bar{T})^{2}}+\ldots\right)\;\;{\rm with}\;A\sim
C_{\alpha'}\;,\;B\sim C_{\rm 1-loop}\label{Vbhk}\eeq generates an
AdS minimum for the volume K\"ahler modulus at \beq {\rm
Re}\,T_0=t_0\sim\frac{1}{\xi^2}\cdot\frac{{\rm Re}\,U^4}{{\rm
Re}\,S^3}\gg 1\eeq if $A<0$, $B>0$ and $B\gg|A|$.

The model leaves - through the omission of simple non-perturbative
effects - unbroken a certain shift symmetry $T\to T+i \alpha$ of
the underlying string theory. The fact that this shift symmetry
remains an unbroken isometry of the supergravity allows for
gauging the shift symmetry into a full nonlinear $U(1)$ gauge
symmetry by turning on magnetic world volume flux on a 4-cycle
wrapping single $D7$-brane. The non-vanishing field dependent
D-term generated this way satisfies all known symmetry
requirements~\cite{BDKP,DuVe,VZ} and can thus provide a consistent
D-term uplift in this model~\cite{PW} (besides the obvious
possibility of inserting an uplifting $\overline{D3}$-brane).

Again, the barrier width of the uplifted resulting dS vacuum can
be determined by extracting the value $t_1>t_0$ satisfying
eq.~\eqref{T1}. Since $V_{\rm AdS}$ approaches zero from below for
$t>t_0$ we can write \beq V_{\rm AdS}\sim
\frac{|W_0|^2}{8\,t^3}\cdot\frac{A}{2\sqrt{2}\,t^{3/2}}\;\;{\rm
for}\;t>t_0\;\;.\label{Vbhk2}\eeq Using then eq.~\eqref{T1} we
determine \beq t_1=t_0\cdot\left[1+\frac{2}{9}\gamma+{\cal
O}(\gamma^2)\right]\;\;{\rm
where}\;\gamma=1-\epsilon\;\;.\label{T1pw}\eeq This, in turn,
implies via eq.~\eqref{width2} a barrier width in terms of the
canonically normalized 'tunneling' field $\phi$ given by 
\bea \Delta\phi_{\rm pert.}&\sim&
\sqrt{6}\cdot\ln\hspace{-0.2ex}\left(\frac{t_1}{t_0}\right)\sim
\left(\frac{2}{3}\right)^{3/2}\cdot(1-\epsilon)=0.05\ldots0.2\;\;{\rm
for}\;\epsilon=0.1\ldots0.2\nonumber\\
&\phantom{\sim}&\;\Rightarrow (\Delta\phi_{\rm pert.})^2\sim10^{-3}\ldots 10^{-2}\gg\frac{V_0}{V_1} \label{PWwidth}\eea
This result
again is parametrically similar to the two other model classes and
thus here, too, a similar estimate results for lifetime of the dS
vacua in these large-volume vacua \beq \tau_{\rm
pert.}\sim\Gamma_{\rm pert.}^{-1}\sim
e^{\frac{24\pi^2}{V_0}}\,t_P\sim e^{10^{120}}\,t_P\;\;{\rm
for}\;V_0\sim 10^{-120}\;\;.\eeq

\subsection{dS vacua in M-theory on $G_2$-manifolds}\label{MdS}

The construction of M-theory on $G_2$-manifolds is described in 4d by the ${\cal N}=1$
chiral ungauged supergravity specified by the K\"ahler potential
and superpotential of~\cite{acharya1,acharya2}
\bea
K=-3 \ln(4\,\pi^{1/3} V_X)+\bar\phi\phi\nonumber\\
W=A_1\phi^{-a}e^{-a_1f^{(1)}(T_i)}+A_2e^{-a_2f^{(2)}(M_i)}\;\;.\label{G2KW}\eea
Here $f^{(1,2)}(T_i)$ denote the gauge kinetic function of the two condensing gauge groups with beta function coefficients $a_1^{-1}\simeq a_2^{-1}$ which in M-theory depend generically on all the moduli fields $T_i=t_i+i\tau_i$ ($\tau_i$ denote the axions). $\phi=(Q\tilde Q)^{(1/2)}$ denotes the meson field formed by a single flavor and anti-flavor of chiral quarks charged under the first gauge group. Its exponent $a$ in the superpotential is given by $a=2/(2\pi a_1^{-1}-1)$. The presence of the meson field allows for the existence of a tunable dS vacuum~\cite{acharya2}. $V_X(T_i)$ denotes the volume of the 7-dimensional $G_2$-manifold $X$. The resulting structure of the scalar potential around the dS minimum for an example case of 2 moduli $T_1,T_2$ with input parameters as in eq.~(137) of~\cite{acharya2} is displayed in Fig.~\ref{fig.dSM}. Note, that the same scalar potential also stabilizes the meson field in a unique minimum implying that here there is no possibility for vacuum decay in the $\phi$--direction.

\FIGURE[ht]{\epsfig{file=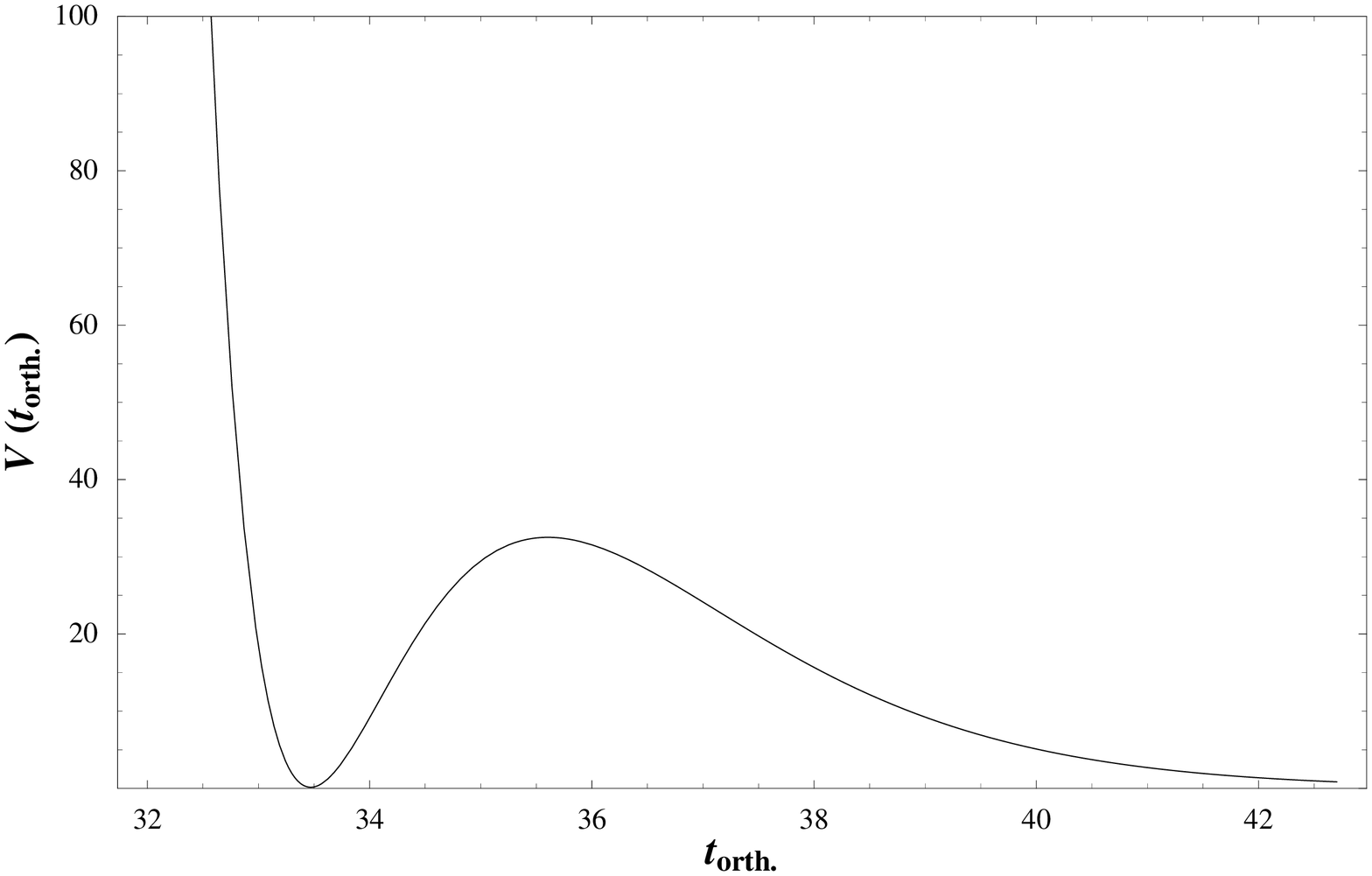,width=12cm} \caption{Black: The F-term scalar
potential $V(t_1,t_2)$ in units of $m_{3/2}^2M_P^2$ in the full case including the stabilized meson field along the 'orthogonal' direction in field space which runs orthogonal to the potential barrier separating the dS minimum from the Minkowski one at infinite volume.}%
\label{fig.dSM}}

In a slight modification of the method
outlined in the Section~\ref{barrierwidthest} we determine the barrier width of the
positive semi-definite scalar potential in Fig.~\ref{fig.dSM} by
extracting the position $t_{{\rm orth.},0}$ of the dS minimum (which is nearly identical with the position of the SUSY AdS saddle point that would be there without the quark flavor) and the
position $t_{{\rm orth.},1}$ of the barrier maximum. The former is still given by the
solution to

\beq \left.D_{T_i}W\right|_{T_{i,0}}\simeq0\;.\label{T0dsM}\eeq

Now from here we can infer the position $t_{{\rm orth.},1}$ of barrier maximum
directly: At $t_{{\rm orth.},1}=t_{{\rm orth.},0}+1/\bar a$ ($\bar a=(a_1+a_2)/2$) the cancellation of the two exponentials in $D_{T_i}W$ is ruined
by a relative factor of $\sim 1/e$ between the exponentials. Again in about twice the distance in $T_i$-field space the potential has decayed from the barrier down to nearly zero giving a barrier width similar to the KKLT case to $\Delta t_{\rm orth.}\sim 2/\bar a$.

Using eq.~\eqref{width2} this gives us the
barrier width in terms of canonically normalized field $\phi$ as

\beq \Delta\phi_{\rm M-theory}\sim
\sqrt{6}\cdot\ln\left(\frac{t_{{\rm orth.},0}+2/\bar a}{t_{{\rm orth.},0}}\right)=\frac{2\sqrt{6}}{\bar a\,t_{{\rm orth.},0}}+{\cal
O}\left(\frac{1}{(\bar a\,t_{{\rm orth.},0})^2}\right)\;.\label{dsMwidth}\eeq
The validity of the supergravity approximation requires $t_{{\rm orth.},0}\gg 1$ and
$\bar a\,t_{{\rm orth.},0}\gg 1$ - typical model constructions like the one shown above have $t_{{\rm orth.},0}={\cal
O}(100)$ and $\bar a\,t_{{\rm orth.},0}={\cal O}(10)$. Thus, in the M--theory model of~\cite{acharya1,acharya2} we
have typically \beq (\Delta\phi_{\rm M-theory})^2=0.01\ldots
0.1\gg\frac{V_0}{V_1}\label{dsMres}\eeq in accordance with the
requirement of eq.~\eqref{gravcrit}. This in turn implies that the
lifetime of the dS vacuum is given from eq.~\eqref{Gamma3}
as \beq \tau_{\rm M-theory}\sim\Gamma_{\rm M-theory}^{-1}\sim
e^{\frac{24\pi^2}{V_0}}\,t_P\sim e^{10^{120}}\,t_P\;\;{\rm
for}\;V_0\sim 10^{-120}\eeq where $t_P\sim 10^{-43}\,s$ denotes
the Planck time.

Finally, as discussed in the Introduction, the life-time estimate we just got explicitly for the above $G_2$-model of~\cite{acharya1,acharya2} will carry over to the construction of strongly coupled heterotic M-theory on $CY_3\times S^1/Z_2$~\cite{CuKr} as there the universal moduli are stabilized by non-perturbative effects alone, too, while all other moduli are fixed at string-scale masses by fluxes.

\section{Conclusion}

In summary, all five model classes lead to metastable dS vacua
with similar barrier width of $\Delta\phi={\cal O}(0.1)$ which
provides all of them with exponentially long vacuum lifetimes
$ \tau\sim\Gamma^{-1}\sim
e^{24\pi^2/V_0}\,t_P\sim e^{10^{120}}\,t_P\;\;{\rm
for}\;V_0\sim 10^{-120}$as long as the barrier potential $V_1$
remains sufficiently large compared with the dS vacuum energy
$V_0$.

The physical reason for this is that the dS constructions of all three classes satisfy a condition on the area of the potential barrier which separates the dS minimum from the Minkowski vacuum at infinity. This condition states according to~\eqref{gravcrit} that the barrier area $O$ must be much larger than the geometric mean $\sqrt{V_0V_1}$ of the potential values of the dS minimum $V_0$ and the barrier $V_1$, that is \[O=\int_{\phi_0}^{\phi_{\infty}}d\phi\,\,V(\phi)\sim V_1\Delta\phi \gg \sqrt{V_0V_1}\;\;\;.\] Since realistic SUSY breaking requires $V_1\gsim (1 {\rm TeV})^4\sim 10^{-60}M_P^4$ and realistic cosmology demands $V_0\sim 10^{-120}M_P^4$ we need \[ O\gg 10^{-90}\sqrt{\frac{V_1[{\rm TeV}^4]}{{\rm TeV}^4}}\] in Planck units to satisfy the above area condition.

But then evaluating the area of the barrier to yield \[{\cal O}\sim  10^{-60}\cdot V_1[{\rm TeV}^4]\cdot\Delta\phi[M_P]\] implies that $V_1\gsim {\rm TeV}^4$ and $\Delta\phi={\cal O}(10^{-3}\ldots 1)$ constitute a barrier high and thick enough to guarantee the above area condition and thus the validity of the thin-wall approximation and the gravitational suppression of vacuum decay. This, in turn, guarantees the exponential longevity of a vacuum. The past Sections then showed that $\Delta\phi={\cal O}(10^{-3}\ldots 1)$ holds for all the constructions discussed which closes the argument.

Note again that this argument relies on the implicit assumption that the uplifted de Sitter vacua is the only meta--stable minimum besides the Minkowski runaway minimum at infinite volume. If there would have been, e.g., two AdS vacua of comparable but different depth prior to uplifting with the more shallow one at smaller volume, then after uplifting the more shallow one to Minkowski the deeper remains still an AdS minimum. In that situation the gravitational correction can enhance tunneling a lot~\cite{LindeSinks} which is why we studied here the classes of stringy de Sitter vacua where such a structure does not appear.

\acknowledgments

I would like to thank M.~Berg for several intensive
discussions and useful comments.


\begin{thebibliography}{99}


\bibitem{KKLT}
S.~Kachru, R.~Kallosh, A.~Linde \& S.~P.~Trivedi, Phys.\ Rev.\ D
{\bf 68}, 046005 (2003) [arXiv:\hepth{0301240}].

\bibitem{GKP}
S.~B.~Giddings, S.~Kachru \& J.~Polchinski, Phys.\ Rev.\ D {\bf
66}, 106006 (2002) [arXiv:\hepth{0105097}].

\bibitem{G2a}
  B.~S.~Acharya,
  Adv.\ Theor.\ Math.\ Phys.\  {\bf 3}, 227 (1999)
  [arXiv:\hepth{9812205}];\\{}
  B.~S.~Acharya,
  arXiv:\hepth{0011089}.
  
\bibitem{G2b}
  M.~Atiyah \& E.~Witten,
  Adv.\ Theor.\ Math.\ Phys.\  {\bf 6}, 1 (2003)
  [arXiv:\hepth{0107177}].

\bibitem{G2c}
  B.~Acharya \& E.~Witten,
  arXiv:\hepth{0109152}.

\bibitem{BeCuKr}
  M.~Becker, G.~Curio \& A.~Krause,
  Nucl.\ Phys.\  B {\bf 693}, 223 (2004)
  [arXiv:\hepth{0403027}].
  
\bibitem{CuKr}
  G.~Curio \& A.~Krause,
  Phys.\ Rev.\  D {\bf 75}, 126003 (2007)
  [arXiv:\hepth{0606243}].

\bibitem{GVW}
S.~Gukov, C.~Vafa \& E.~Witten, Nucl.\ Phys.\ B {\bf 584}, 69
(2000) [Erratum-ibid.\ B {\bf 608}, 477 (2001)]
[arXiv:\hepth{9906070}].

\bibitem{acharya1}
  B.~Acharya, K.~Bobkov, G.~Kane, P.~Kumar \& D.~Vaman,
  Phys.\ Rev.\ Lett.\  {\bf 97}, 191601 (2006)
  [arXiv:\hepth{0606262}].

\bibitem{acharya2}
  B.~S.~Acharya, K.~Bobkov, G.~L.~Kane, P.~Kumar \& J.~Shao,
  arXiv:\hepth{0701034v3}.

\bibitem{BKQ}
C.~P.~Burgess, R.~Kallosh \& F.~Quevedo, JHEP {\bf 0310}, 056
(2003) [arXiv:\hepth{0309187}].

\bibitem{Carlos}
A.~Achucarro, B.~de Carlos, J.~A.~Casas \& L.~Doplicher,
[arXiv:\hepth{0601190}].

\bibitem{DuMamb}
E.~Dudas \& Y.~Mambrini, JHEP {\bf 0610}, 044 (2006)
[arXiv:\hepth{0607077}].

\bibitem{HaackKrefl}
M.~Haack, D.~Krefl, D.~Lust, A.~Van Proeyen \& M.~Zagermann,\\{}[arXiv:\hepth{0609211}].

\bibitem{Qbkq2}
D.~Cremades, M.~P.~G.~del Moral, F.~Quevedo \& K.~Suruliz, [arXiv:\hepth{0701154}].



\bibitem{BurgessD3}
C.~P.~Burgess, J.~M.~Cline, K.~Dasgupta \& H.~Firouzjahi, [arXiv:\hepth{0610320}].

\bibitem{nilles}
O.~Lebedev, H.~P.~Nilles \& M.~Ratz, Phys.\ Lett.\  B {\bf 636}, 126 (2006)
[arXiv:\hepth{0603047}].

\bibitem{ReinoScrucca}
M.~Gomez-Reino \& C.~A.~Scrucca, JHEP {\bf 0605}, 015 (2006)
[arXiv:\hepth{0602246}].

\bibitem{DuISS}
E.~Dudas, C.~Papineau \& S.~Pokorski, [arXiv:\hepth{0610297}].

\bibitem{AHKO}
H.~Abe, T.~Higaki, T.~Kobayashi \& Y.~Omura,
[arXiv:\hepth{0611024}]

\bibitem{OKKLT1}
F.~Br\"ummer, A.~Hebecker \& M.~Trapletti, Nucl.\ Phys.\ B {\bf
755}, 186 (2006) [arXiv:\hepth{0605232}].

\bibitem{OKKLT2}
R.~Kallosh \& A.~Linde, [arXiv:\hepth{0611183}].

\bibitem{ISS}
K.~Intriligator, N.~Seiberg \& D.~Shih, JHEP {\bf 0604}, 021
(2006) [arXiv:\hepth{0602239}].

\bibitem{bbhl}
K.~Becker, M.~Becker, M.~Haack \& J.~Louis, JHEP {\bf 0206}, 060
(2002) [arXiv:\hepth{0204254}].

\bibitem{Brama}
V.~Balasubramanian \& P.~Berglund, JHEP {\bf 0411}, 085 (2004)
[arXiv:\hepth{0408054}].

\bibitem{West}
A.~Westphal, JCAP {\bf 0511}, 003 (2005) [arXiv:\hepth{0507079}].

\bibitem{West2}
A.~Westphal, JHEP {\bf 0703}, 102 (2007) [arXiv:\hepth{0611332}].


\bibitem{BBCQ}
V.~Balasubramanian, P.~Berglund, J.~P.~Conlon \& F.~Quevedo, JHEP
{\bf 0503}, 007 (2005) [arXiv:\hepth{0502058}].

\bibitem{CQS}
J.~P.~Conlon, F.~Quevedo \& K.~Suruliz, JHEP {\bf 0508}, 007
(2005) [arXiv:\hepth{0505076}].


\bibitem{Silverstein}
A.~Saltman \& E.~Silverstein, JHEP {\bf 0601}, 139 (2006)
[arXiv:\hepth{0411271}].

\bibitem{Silverstein2}
A.~Maloney, E.~Silverstein \& A.~Strominger, [arXiv:\hepth{0205316}];\\{}


\bibitem{Moral}
M.~P.~Garcia del Moral, JHEP {\bf 0604}, 022 (2006) [arXiv:\hepth{0506116}].

\bibitem{BHK1}
M.~Berg, M.~Haack \& B.~Kors, JHEP {\bf 0511}, 030 (2005)
[arXiv:\hepth{0508043}].

\bibitem{GH}
G.~von Gersdorff \& A.~Hebecker, Phys.\ Lett.\ B {\bf 624}, 270
(2005) [arXiv:\hepth{0507131}].

\bibitem{BHK2}
M.~Berg, M.~Haack \& B.~Kors, Phys.\ Rev.\ Lett.\  {\bf 96},
021601 (2006) [arXiv:\hepth{0508171}].

\bibitem{PW}
S.~L.~Parameswaran \& A.~Westphal, JHEP {\bf 0610}, 079 (2006) [arXiv:\hepth{0602253}].

\bibitem{CDL}
S.~R.~Coleman \& F.~De~Luccia, Phys.\ Rev.\ D {\bf 21}, 3305
(1980).

\bibitem{KL}
R.~Kallosh \& A.~Linde, JHEP {\bf 0412}, 004 (2004) [arXiv:\hepth{0411011}].

\bibitem{LindeSinks}
A.~Ceresole, G.~Dall'Agata, A.~Giryavets, R.~Kallosh \& A.~Linde, Phys.\ Rev.\ D {\bf 74}, 086010 (2006) [arXiv:\hepth{0605266}].

\bibitem{LVSloops}
  M.~Berg, M.~Haack \& E.~Pajer,
  arXiv:0704.0737 [hep-th].


\bibitem{BDKP}
P.~Binetruy, G.~Dvali, R.~Kallosh \& A.~Van Proeyen, Class.\
Quant.\ Grav.\  {\bf 21}, 3137 (2004) [arXiv:\hepth{0402046}].

\bibitem{DuVe}
E.~Dudas \& S.~K.~Vempati, Nucl.\ Phys.\ B {\bf 727}, 139 (2005)
[arXiv:\hepth{0506172}].

\bibitem{VZ}
G.~Villadoro \& F.~Zwirner, Phys.\ Rev.\ Lett.\  {\bf 95}, 231602
(2005) [arXiv:\hepth{0508167}].



\end{thebibliography}
\end{document}